# Experimental tests for macroscopic phase coherence in magnetic-quasiparticle condensates of insulating spin systems

A Schilling, H Grundmann[1] and R Dell'Amore[1]
Physik-Institut, University of Zurich, Winterthurerstrasse 190,
CH-8057 Zurich, Switzerland

E-mail: schilling@physik.uzh.ch

**Abstract.** For most kinds of already known Bose-Einstein condensates experimental evidence for the existence of a macroscopic coherent quantum state has been provided, e.g., from the observation of interference phenomena, the formation vortices, the detection of Josephson effects, or even by the manifestation of superfluid behaviour. However, none of these hallmarks for a true, macroscopic phase-coherent state has ever been reported for any insulating spin system in a solid in which magnetic bosonic quasiparticles are thought to condense close to a quantum critical point. We describe an attempt to detect superfluid behaviour in $TlCuCl_3$, and discuss in more detail a proposal for an experiment in which the a.c. Josephson effect can be probed by using a device composed of two weakly coupled magnetic insulators with different critical fields. We conclude that the detection of the a.c. Josephson effect is feasible for a proper choice of compounds with realistic material parameters.

## 1. Introduction

Quantum spin systems in solids are a subject of intense research, both theoretically and experimentally. Certain insulating magnetic systems show field-induced phase transitions at zero temperature that are commonly interpreted as a Bose-Einstein condensation (BEC) of magnetic bosonic quasiparticles [1,2]. The stability of such condensates [3,4] and even the applicability of the BEC concept as a whole [5] have been questioned, however. Already known Bose-Einstein condensates include superfluid helium, dilute atomic gas clouds and pumped exciton-polariton and magnon condensates. Several experimental proofs for the existence of a macroscopic quantum state have been reported for the latter condensates, e.g., from interference experiments, from the observation of vortices and Josephson effects, and even superfluid properties can be ascribed to most of these systems. Such hallmarks for a true BEC state are all based on the existence of a macroscopic phase-coherent state, but none of these has ever been observed in any insulating quantum spin system.

    In the following we summarize some assumptions that are commonly made when treating the antiferromagnetic state of certain insulating spin systems as a condensate of magnetic bosonic quasiparticles. We then discuss the resulting consequences for experiments that aim to probe macroscopic phase coherence, and we finally propose an experiment that could provide a decisive evidence for the existence of a macroscopic phase-coherent condensate in such systems.

[1] Swiss National Foundation grants 20-111653 and 21-126411

## 2. Background

The most frequently discussed magnetic systems in the context of a BEC of magnetic bosonic quasiparticles in solids are insulating dimerized spin-1/2-systems [1,2]. In a perfectly axially symmetric case (with the symmetry axis parallel to the external magnetic field *H*) one expects a condensation of triplet bosonic quasiparticles ("triplons") above a certain critical field $H_{c1}$ where the energy difference between the ground-state singlet and the lowest excited triplet states vanishes due to the Zeeman splitting. The resulting antiferromagnetic state is characterized by the presence of a staggered transverse magnetic moment perpendicular to the external magnetic field, the absolute value and relative angle of which to the crystal lattice are related to the amplitude and to the phase of the condensate wave function, respectively, and the longitudinal component is proportional to the number of condensed bosons [1,2]. This approach has indeed lead to a number of successful predictions, such as for the details of the magnetic phase diagram of the antiferromagnetic phase [2], for the respective magnon dispersion [6] and for the magnetization components parallel and perpendicular to the external magnetic field [2]. The measurement of a characteristic linear energy-versus-momentum $E(k)$ relation in the magnetic excitation spectrum of the spin-dimer system $TlCuCl_3$ down to energies of the order of $\approx 0.75$ meV [6] has very early been taken as a proof for the existence of a gapless Goldstone mode, which is an essential prerequisite for invoking BEC physics.

### 2.1. *Triplons as a weakly interacting Bose gas*

As a consequence of a weak coupling between otherwise strongly coupled dimers, magnetic excitations show dispersion, the bandwidth of which is essentially determined by the coupling between the dimers [1]. This bandwidth (together with the intra-dimer coupling energy) not only determines the values of the critical fields $H_{c1}$ beyond which antiferromagnetic order develops and the saturation field $H_{c2}$ above which all spins can be thought to be aligned in parallel, it also correlates with the parameter $v_0$ which is a measure for the interaction energy between the "hard-core" bosons that are supposed to condense beyond $H_{c1}$ (in the following we identify $H_{c1}$ with the "critical field $H_c$"). In contrast to the case of a condensate of non-interacting bosons ($v_0 = 0$), the chemical potential $\mu$ (i.e., the energy gain/loss upon adding/removing one particle to/from the condensate) of a BEC of weakly interacting bosons is non-zero, and it can be expressed by the fraction *n* of condensed particles according to $\mu = n v_0$ [1,2]. In the present case, $\mu$ (and along with it *n*) is determined by *H* according to

$$\mu = g\mu_B\mu_0(H - H_c),  \qquad (1)$$

where *g* is the Landé *g*-factor and $\mu_B$ the Bohr magneton.

The macroscopic wave function $\psi_0(r,t) = |\psi_0| e^{i\phi(r,t)}$ describing the resulting condensate minimizes a potential energy functional [3], namely

$$u(\psi) = -\mu \psi^*\psi + \frac{v_0}{2}(\psi^*\psi)^2. \qquad (2)$$

For sufficiently dilute, light bosons, the number of non-condensed quasiparticles is negligibly small and the resulting condensate fraction at $T = 0$ is $n = \psi_0^*\psi_0$, here defined as the total number of condensed triplons $N(0)$ divided by the number of dimers $N_d$.

A necessary requirement for the conservation of the number of quasiparticles is the rotational invariance (axial symmetry) of the system with respect to the direction of the magnetic field, so that the Hamiltonian describing the bosonic system commutes with the corresponding particle number operator. In this case, the phase $\phi$ is completely arbitrary and can accommodate any value.

## 2.2. Role of amplitude and phase of the macroscopic wave function

The quantity $n = \psi_0 * \psi_0 = |\psi_0|^2$ is proportional to the longitudinal component of the magnetic moment of the condensate, $M_z = g\mu_B n N_d$ [2].

The role of the phase of the macroscopic wave function in magnetic insulators is played by the angle $\varphi$ of the staggered transverse magnetic moments within the plane perpendicular to the main magnetic field $H$ [1]. A plausibility argument for this fact is that the particle number operator is canonically conjugate to the phase $\phi$ of the macroscopic wave function on the one hand, but also (via the proportionality of $N = nN_d$ to $M_z$ and to the associated component of the angular momentum $L_z$) to the angle variable $\varphi$ in the plane perpendicular to $H$ on the other hand. As a consequence of the arbitrariness of the phase $\phi$, the in-plane angle $\varphi$ of the transverse magnetic moments in the condensed state must be able to take any value between 0 and $2\pi$, which is possible only in an axially symmetric spin system. The condensate may also support a "spin-supercurrent", the maximum velocity of which is given by the Landau criterion,

$$v = \hbar/m^* |\vec{\nabla}\phi| < v_c \equiv \min(E(k)/\hbar k) = (nv_0/m^*)^{1/2}, \qquad (3)$$

where $m^*$ is the effective mass of the quasiparticles. Such a supercurrent would be associated with a finite gradient of the angles of the transverse moments along a direction perpendicular to $H$.

## 3. Superfluidity in TlCuCl$_3$?

An early proposal to probe a superfluid property of a triplon BEC was based on the argument that the occurrence of a superfluid BEC might manifest itself in a dramatically enhanced thermal conductivity as it is observed in superfluid helium, but corresponding measurements on TlCuCl$_3$ ($\mu_0 H_c \approx 5.5$ T [2]) showed only a moderate increase in the total thermal conductivity [7]. It is conceivable that the coupling of the spin system to the monoclinic crystal lattice of TlCuCl$_3$ is either too strong on the one hand, so that the condition of axial symmetry is not fulfilled (see also section 4.1), or too weak on the other hand, so that external heat sources and macroscopic thermometers essentially probe the properties of the lattice in such an experiment.

As an alternative to applying a directed heat current, we focused on the idea that sufficiently long-lived collective modes in a superfluid BEC state should be excitable, that can be regarded as analogues to standing entropy or temperature waves ("second sound") observed in superfluid helium [8]. As the quasiparticles are of magnetic origin, the necessary excitation could be provided by an external a.c. magnetic field. To estimate at which frequencies such standing waves can be expected we have modelled the relevant modes as three dimensional standing waves in a medium with anisotropic group velocities ($c_a \approx 380$ m/s, $c_b \approx 2400$ m/s, $c_c \approx 2600$ m/s, estimated from published calculated magnon-dispersion $E(k)$ relations), and appropriate boundary conditions according to the sample geometry (an almost rectangular solid with dimensions $\approx$ 2x2x7 mm$^3$) [8]. Within this model, the first resonance should appear at $f \approx 470$ kHz, and a characteristic sequence of higher-order resonances should follow at higher excitation frequencies. We have therefore built a series of FM modulated a.c. susceptometers to perform high frequency (up to $f = 2$ MHz) magnetic-susceptibility measurements at $T = 2$ K and in $\mu_0 H = 9$ T with a sensitivity of the order of 2 x 10$^{-13}$ Am$^2$ (2 x 10$^{-10}$ emu), which is almost an order of magnitude better than what is reached in commercial SQUID magnetometers. However, based on these experiments and within our sensitivity limit, we can exclude corresponding resonances in TlCuCl$_3$ on a time scale larger than $\tau \approx 0.5\ \mu$s [8].

## 4. The a.c. Josephson effect

We now consider a system of two dimerized spin systems ($\alpha$ and $\beta$) at zero temperature and with different critical fields $H_{c\alpha} \neq H_{c\beta}$ beyond which the respective magnetic quasiparticles are supposed to condense, with $g_\alpha = g_\beta = g$ for simplicity. The boundary layers are assumed to weakly couple to one another to allow for a small but finite magnetic coupling and therefore for a tunneling of magnetic quasiparticles, and we place the device into an external magnetic field $H$ (see figure 1). As a result of the difference between the critical fields, the magnetic field $H > H_{c,j}$ ($j = \alpha$ or $\beta$) maintains a constant difference $\Delta\mu = \mu_\alpha - \mu_\beta$ between the respective chemical potentials (see figure 1a), in analogy to the external voltage that controls $\Delta\mu$ in superconducting Josephson junctions. This non-zero difference is not simply a result of choosing different reference points of the energy scale for the two condensates. The difference in the critical fields $H_{c,j}$ originally stems from different microscopic inter- and intradimer coupling constants that determine the individual energy gaps, i.e. the differences between the energies of the respective ground-state singlet and the triplet states in zero magnetic field. To close these energy gaps, different external magnetic fields have to be applied beyond which the respective ground-state triplons condense (see inset of figure 2). Therefore the chemical potentials $\mu_j$ are different in a common external magnetic field $H$ exceeding both $H_{c\alpha}$ and $H_{c\beta}$ (see figure 1a).

In an approach introduced by Feynman to explain the Josephson effects across a weak link between two superconductors one considers the macroscopic wave functions $\psi_j = n_j^{1/2} e^{i\phi_j}$ on opposite sides of the junction, and treats the weak coupling between them according to $i\hbar\partial\psi_{\alpha;\beta}/\partial t = \mu_{\alpha;\beta}\psi_{\alpha;\beta} + K\psi_{\beta;\alpha}$, where $K$ is a phenomenological coupling constant. The resulting differential equations for the number of particles $n_j$ occupying the respective macroscopic quantum states and for the corresponding phase difference $\Delta\phi = \phi_\beta - \phi_\alpha$ become

$$\frac{\partial n_\alpha}{\partial t} = -\frac{\partial n_\beta}{\partial t} = \frac{2K}{\hbar}\sqrt{n_\alpha n_\beta}\sin(\Delta\phi), \qquad (4)$$

$$\frac{\partial\Delta\phi}{\partial t} = \frac{\Delta\mu}{\hbar} - \frac{K}{\hbar}\frac{(n_\alpha - n_\beta)}{\sqrt{n_\alpha n_\beta}}\cos(\Delta\phi). \qquad (5)$$

As the main magnetic field penetrates the whole sample volume, thereby keeping $\Delta\mu$ constant and acting as a source or sink of magnetic quasiparticles (in a similar way to an external battery in superconducting Josephson junctions), there is no macroscopic directional flow of quasiparticles within the two branches of the junction, except in direct proximity to the spacer region forming the weak link [9]. Therefore, the variation $\partial n_j/\partial t$ from Eq. (4) may be unobservable.

For $K \neq 0$ and $K \ll \Delta\mu$, equation 5 denotes a narrowband frequency modulation around the carrier frequencies $\mu_j/\hbar$ with a modulation frequency $\omega_{a.c.}$ and a modulation index $K/\Delta\mu$ [9], with

$$\phi_{\alpha;\beta}(t) \approx \phi_{0,\alpha;\beta} - \frac{\mu_{\alpha;\beta}}{\hbar}t - \frac{K}{\Delta\mu}\sqrt{\frac{n_{\beta;\alpha}}{n_{\alpha;\beta}}}\sin(\omega_{a.c.}t + \Delta\phi_0), \text{ where } \Delta\phi_0 = \phi_{0,\beta} - \phi_{0,\alpha}. \qquad (6)$$

As a result, equally spaced sidebands should appear in the energy spectrum of the coupled device (see figure 1c). Such sidebands are absent for two uncoupled magnetic insulators (figure 1b) and, of course, also in a scenario where no macroscopic phase coherence is present at all. Therefore, a successful experiment in a coupled system that can probe the occurrence of sidebands with separation $\hbar\omega_{a.c.}$ would represent a very strong experimental support for the existence of a state with macroscopic phase coherence. A high-resolution electron-spin resonance (ESR) measurement of the transitions between the (condensed) ground state considered here, and the nearest excited, (uncondensed) triplet states (referred to as $A_0$ in [10]) should reveal a characteristic splitting of the corresponding modes by $\hbar\omega_{a.c.}$ which is related to the a.c. Josephson effect. As an illustration, we show in figure 2a sketch of an expected ESR signal based on a corresponding measurement on TlCuCl$_3$ taken from [10].

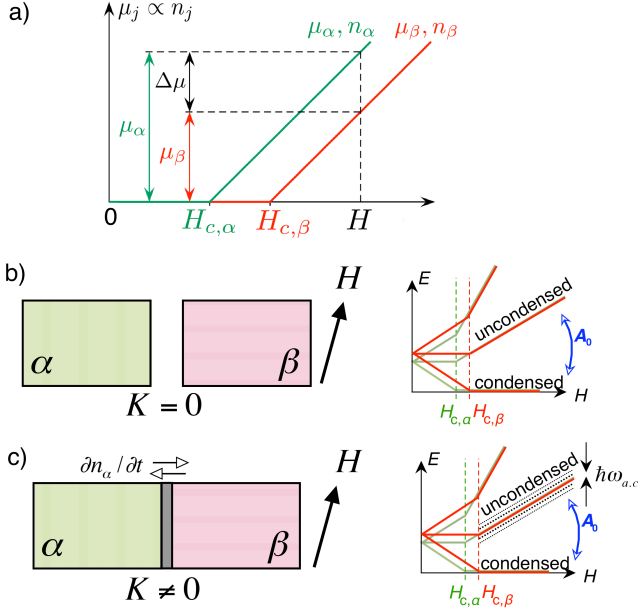

**Figure 1.** Sketch of an experiment in which two magnetic insulators ($j = \alpha$ and $\beta$) with different critical fields $H_{c,j}$ are weakly coupled to one another by a magnetically inert spacer layer, and are placed into an external magnetic field $H > H_{c,j}$.

a) The respective fractions of condensed quasiparticles $n_j$ and the chemical potentials $\mu_j$ as functions of $H$.

b) and c): Sketch of the energy schemes for b) two uncoupled magnetic insulators with different critical fields, and c) two weakly coupled magnetic insulators (a.c. Josephson effect, frequency modulation with $\omega_{a.c.} = \Delta\mu/\hbar$, see text).

4.1. *Necessary conditions and constraints*

In order to be able to resolve the ESR sidebands conjectured above, the width of these bands (which is related to the lifetime $\tau_{pc}$ of phase coherence) must be smaller than their spacing, i.e., $\tau_{pc} > \omega_{a.c.}^{-1}$. This may already be the case if the lifetime of the condensed magnetic quasiparticles $\tau_{qp}$ fulfils $\tau_{qp} > \omega_{a.c.}^{-1}$ because we expect $\tau_{pc} \geq \tau_{qp}$ [9]. The $\tau_{qp}$ can be estimated, for example, from the width of corresponding peaks in inelastic neutron-scattering or ESR experiments.

A more serious problem may be the occurrence of an anisotropy gap in $E(k)$ that is lifting the Goldstone linearity for $k \to 0$. This field-dependent gap has been estimated to $\Delta_a = (8\tilde{\gamma}\mu)^{1/2}$ [11], where $\tilde{\gamma}$ denotes an exchange-anisotropy parameter that is characteristic for a given material. In TlCuCl$_3$, $\tilde{\gamma} \approx 16$-$30$ $\mu$eV [3], while $\tilde{\gamma} \approx 1$ $\mu$eV in Ba$_3$Cr$_2$O$_8$ or BaCuSi$_2$O$_6$ [9]. We cannot expect phase coherence on time scales longer than $\tau_\Delta = \hbar/\Delta_a$, or for too small experimental frequencies $\omega < \tau_\Delta^{-1}$, respectively [3,8]. This fact may be a conclusive explanation why our early experiments with a $\omega \approx 1.2 \times 10^7$ s$^{-1}$ on TlCuCl$_3$ failed, for which we estimate $\tau_\Delta \approx 3 \times 10^{-12}$ s (see section 3).

Finally, the width of the spacer region forming the weak link should not be much larger than the healing length $\xi = \hbar/(2nv_0 m^*)^{1/2}$, the analogue to the coherence length in superconductors. The latter condition can be enforced by operating the experiment near the Bose-Einstein condensation temperature where $n$ vanishes, but probably at the cost of reducing $\tau_{qp}$ and $\tau_{pc}$.

4.2. *(Ba,Sr)$_3$Cr$_2$O$_8$: A toy model system*

The isostructural compounds Ba$_3$Cr$_2$O$_8$ and Sr$_3$Cr$_2$O$_8$ are known to have critical fields $\mu_0 H_c \approx 12$ T and 30 T respectively, and we can hypothesize that a mixed composition (Ba,Sr)$_3$Cr$_2$O$_8$ exists with a $\mu_0 H_c \approx 12.5$ T [9]. For a junction composed of Ba$_3$Cr$_2$O$_8$ ($\alpha$) and (Ba,Sr)$_3$Cr$_2$O$_8$ ($\beta$) we obtain with $g \approx 2$ $\omega_{a.c.} \approx 10^{11}$ s$^{-1}$. The corresponding quasiparticle lifetime in dimerized spin systems can be estimated to up to $\tau_{qp} \approx 5 \times 10^{-11}$ s $> \omega_{a.c.}^{-1}$ [9]. In an external magnetic field $\mu_0 H = 13$ T and with $\tilde{\gamma} \approx 1$ $\mu$eV we also have $\tau_\Delta \approx 2 \times 10^{-11}$ s $> \omega_{a.c.}^{-1}$. Using $m^* \approx 1.5 \times 10^{-27}$ kg and with $v_0/k_B \approx 8.7$ K, we finally obtain with $n_\alpha \approx 0.15$ and $n_\beta \approx 0.08$ $\xi_j \approx 0.4$-$0.6$ nm, i.e., a tiny gap or one spacer layer of nonmagnetic isostructural Ba$_3$V$_2$O$_8$ should be adequate to form the necessary weak link [9].

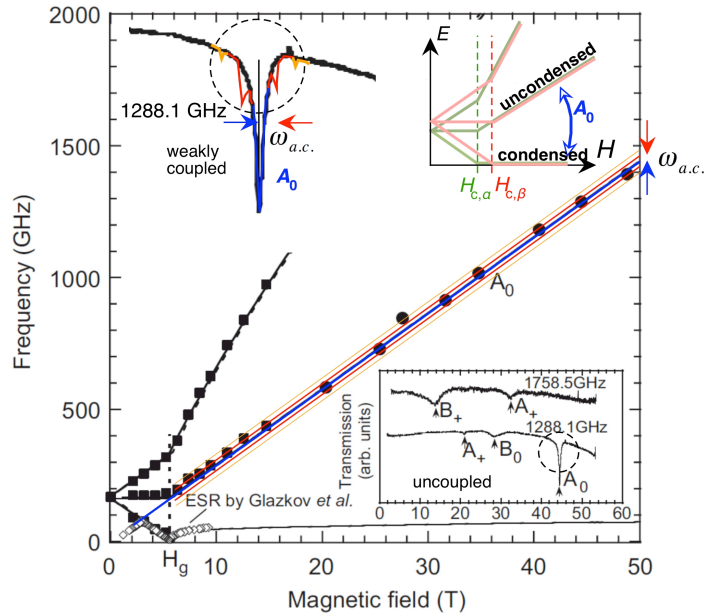

**Figure 2.** Sketch of an ESR spectrum as we expect it in an experiment probing the a.c. Josephson effect, based on ESR data for TlCuCl$_3$ that we reproduced from [10]. Only the "condensed" triplon branch is affected by the frequency modulation (top right inset), which results in a splitting of the $A_0$ mode by $\omega_{a.c.}$ (top left inset and main panel).

## 5. Conclusions

We have addressed the question whether the scenario of a Bose-Einstein condensation of magnetic bosonic quasiparticles results in a state with experimentally testable macroscopic phase coherence. We conclude that two weakly linked magnetic insulators with different critical fields and with suitable material parameters should make the generation and the observation of the a.c. Josephson effect in magnetic insulators feasible in principle. It could be probed, for example, by electron-spin resonance methods.